\documentclass[aps,prd,preprint,groupedaddress,showpacs,showkeys,nofootinbib]{revtex4-1}

\usepackage{amsmath,amssymb,MnSymbol}
\usepackage{graphicx}
\usepackage[colorlinks,linkcolor=red,anchorcolor=blue,citecolor=blue]{hyperref}

\begin{document}
\def\query#1{\marginpar{\begin{flushleft}\footnotesize#1\end{flushleft}}}
\newcommand{\br}{\bra{p',s'}}
\newcommand{\kt}{\ket{p,s}}
\newcommand{\p}{\partial_}
\newcommand{\m}{\mathbf}{}
\newcommand{\bs}{\boldsymbol}{}
\newcommand{\tf}{\tau_1}
\newcommand{\ts}{\tau_2}
\newcommand{\eq}{\begin{eqnarray}}
\newcommand{\en}{\end{eqnarray}}

\title{Three-particle quantization condition in a finite volume:\\
1. The role of the three-particle force}


\author{Hans-Werner Hammer$^{a}$, Jin-Yi Pang$^{b}$ and Akaki Rusetsky$^{b}$\\~}

\affiliation{$^a$Institut f\"ur Kernphysik, Technische Universit\"at Darmstadt,
  64289 Darmstadt, Germany and \\ ExtreMe Matter Institute EMMI,
  GSI Helmholtzzentrum f\"ur Schwerionenforschung, 64291 Darmstadt, Germany\\
\\
$^b$Helmholtz-Institut f\"ur Strahlen- und Kernphysik (Theorie) and\\
Bethe Center for Theoretical Physics, Universit\"at Bonn,\\ D-53115 Bonn, Germany}

\date{\today}

\begin{abstract}
Using non-relativistic effective Lagrangians in the particle-dimer picture, we rederive
the expression for the energy shift of a loosely bound three-particle bound state
of identical bosons in the unitary 
limit. The effective field theory formalism allows us to investigate the role of the
three-particle force,
which has not been taken into account in the earlier treatment of the problem. 
Moreover, we are able to relax the requirement of the unitary limit of infinite scattering
length and demonstrate a smooth transition from the weakly bound three-particle state to a
two-particle bound state
of a particle and a deeply bound dimer.

\end{abstract}


\maketitle

\setcounter{section}{0}

\section{Introduction}

For some time now lattice QCD calculations have been addressing hadron physics problems
which involve the dynamics of three or more hadrons. As an example,
we quote the calculation of the parameters of the Roper
resonance~\cite{Mathur:2003zf,Guadagnoli:2004wm,Leinweber:2004it,Sasaki:2005ap,Sasaki:2005ug,Burch:2006cc,Liu:2014jua,Mahbub:2010jz,Lasscock:2007ce,Lang:2016hnn},
which decays -- at a substantial rate -- into a nucleon and two pions.
The finite-volume effects in such 
few-body systems are expected to be rather pronounced. Hence,
understanding  these effects is a necessary precondition for investigating 
intriguing puzzles such as the level ordering of the $N^*(1440)$ and the $N^*(1535)$.
Another reason for studying the
three-body dynamics in a finite volume now is the recent advent of
lattice QCD calculations of light nuclei~\cite{Beane:2010em,Beane:2012vq,Chang:2015qxa}
and corresponding calculations in nuclear effective field theory on the 
lattice~\cite{Epelbaum:2009pd,Epelbaum:2011md,Rokash:2013xda,Elhatisari:2015iga}. In order to fully
exploit these advances, a formalism is needed to translate the ``raw'' lattice results into
physical observables like cross sections into various two- and three-body channels,
etc. It is also important that the proposed formalism is not overly complicated and can
be used even when only a few data points are available.
  
The quantization of the energy levels of a three-particle system in a finite volume
has been considered first in Ref.~\cite{Polejaeva:2012ut}. In a series of subsequent papers
by different groups~\cite{Meissner:2014dea,Guo:2016fgl,Guo:2017ism,Briceno:2012rv,Hansen:2014eka,Hansen:2015zga,Hansen:2015zta,Hansen:2016fzj,Hansen:2016ync,Briceno:2017tce} further important aspects of the problem have been addressed.
We would like to stress, however, that despite the substantial progress made, the
formalism is still very complicated and, in our opinion, not quite ready to be 
used straightforwardly by lattice practitioners (in contrast to, e.g., the L\"uscher
equation for two-particle elastic scattering~\cite{Luescher-torus}). In addition,
the relation between the different approaches is not obvious and has not been 
discussed in the literature so far. At the same time,
in Refs.~\cite{Kreuzer:2010ti,Kreuzer:2009jp,Kreuzer:2008bi,Kreuzer:2012sr}, the
volume dependence of the discrete three-body spectrum has been investigated
for bosons as well as nucleons by solving 
the bound state equations in a finite volume numerically. An effective field
theory in the dimer formalism has been used to derive the finite volume bound state
equations and relate the bound state properties to scattering parameters in
the infinite volume, which greatly simplifies the handling of the three-body problem.
These studies also suggest a strategy for formal investigations of
three-body dynamics in a finite volume.

The aim of our work is to provide a simple formalism for the analysis of the
present and forthcoming lattice data in a straightforward and
transparent manner, and to understand the link with the earlier approaches. 
To this end, we use the particle-dimer approach, which 
is very convenient and allows one to achieve this goal with a surprising ease. 
We also found it justified to split the material in two parts. In the first paper, we consider
a very simple system -- a shallow three-body bound state in a finite volume -- which
has been already studied in the literature~\cite{Meissner:2014dea,Hansen:2016ync}.
A leading-order analytic expression for the finite-volume energy shift of this system
in the unitary limit is available, and we shall see how this result can easily be obtained
in the particle-dimer picture. In addition, we address the following issues which
were not considered in previous work: 

\begin{itemize}

\item[(i)]
  The role of the three-body force is studied explicitly. We demonstrate
  that this short-range force does not affect the analytic form of the
  leading-order volume dependence of the shallow bound state energy.
However, it is important to know how the three-body 
force enters in finite-volume observables. Note that omitting the three-body
force altogether renders the three-body problem ill-defined in the infinite volume,
since the whole renormalization program fails 
(see, e.g., Refs.~\cite{Bedaque:1998kg,Bedaque:1998km}).\footnote{The case of a
covariant formulation was recently investigated in Ref.~\cite{Epelbaum:2016ffd}.} 
Hence, its inclusion is matter of principle and does not simply amount to evaluating corrections
to the leading-order result.

\item[(ii)]
  The leading-order result for the finite volume dependence of a shallow
  three-body bound state is derived for finite scattering length.
  From the previous derivation in the unitary limit, it is
not immediately clear how one can move beyond this approximation, as well as
how to build in effective range corrections, mixing with higher partial waves, etc..

\item[(iii)]
The formalism developed in Refs.~\cite{Briceno:2012rv,Hansen:2014eka} explicitly
excludes deeply-bound dimers. It is not obvious how to proceed if
such deeply-bound states exist. We address this issue and show
how the leading-order result for a shallow three-body state
goes over into the well-known result for the particle-dimer bound state that can be
obtained with the use of the L\"uscher equation.   

\end{itemize}

In the present paper, we elaborate on the above issues for a system of three
identical bosons. The study of this 
particularly simple model lays the foundation for
the treatment of the general three-body quantization
condition in the dimer picture, which is considered in our forthcoming 
publication~\cite{forthcoming}.

The layout of the paper is follows. In Section~\ref{sec:twobody}, we consider the role of
short-range interactions in the finite-volume behavior of the two-body binding energy.
This simple example illustrates the pattern, along which the inclusion of the short-range
three-particle force is considered. In Section~\ref{sec:infinite}, we collect all the information
about the particle-dimer formalism in the infinite volume, including the analytic
solution of the problem in the unitary limit. The leading order formula for the shallow
bound-state energy shift in the unitary limit~\cite{Meissner:2014dea} is rederived in 
Section~\ref{sec:finite}. In addition, we discuss the relation of the 
asymptotic normalization coefficient in this formula to the short-range three-body
force. Section~\ref{sec:beyond} deals with the calculation of the energy shift beyond
the unitary limit and the relation to the particle-dimer bound state picture. Finally,
Section~\ref{sec:concl} contains our conclusions.

\section{Two-body bound state}
\label{sec:twobody}

The energy shift of a shallow two-body bound state of identical bosons 
of mass $m$ in a finite volume is given by the L\"uscher formula~\cite{Luescher-1}
\eq
\Delta E_B=-12\kappa_2 |{\cal A}_2|^2\,\frac{\exp(-\kappa_2 L)}{mL}+\cdots\, ,
\en
where 
$L$ is the size of the cubic box, $\kappa_2=\sqrt{mE_B}$ is the two-body binding momentum and ${\cal A}_2$
denotes the two-body asymptotic normalization coefficient (note that we choose a
different definition for ${\cal A}_2$ as in Ref.~\cite{Luescher-1}). 
The latter is defined through 
the behavior of the radial bound-state wave function at large distances:
\eq
\Psi(r)\sim \sqrt{\frac{\kappa_2}{2\pi}}\,{\cal A}_2\frac{\exp(-\kappa_2r)}{r}\, ,\quad\quad
\mbox{as}~r\to\infty\, .
\en
Let us derive the relation of the quantity ${\cal A}_2$ to the parameters characterizing the
short-range interactions. We shall do this using the non-relativistic effective Lagrangians
with a method that closely resembles the one used in the three-particle case.
Dimensional regularization with minimal subtraction is the most convenient choice,
albeit the final results do not depend on the regularization used. This is explicitly
demonstrated in Appendix~\ref{app:cutoff}, where we demonstrate that exactly same results are
obtained by using cutoff regularization. The same applies for dimensonal regularization
with power divergence subtraction. 

In momentum space, the non-relativistic interaction Hamiltonian consists of an
infinite tower of operators with increasing mass dimension. Limiting ourselves
to $O({\bf p}^2)$ and to S-waves only, in momentum space we may write
\eq
H_{int}({\bf p},{\bf q})\doteq\langle {\bf p}|V|{\bf q}\rangle
=2 C_0+{C_2}\,({\bf p}^2+{\bf q}^2)+O({\bf p}^4)\, ,
\label{eq:Hint}
\en
where the couplings $C_0$ and $C_2$ specifying the effective potential $V$
are related to the scattering length $a$
and the effective range $r_e$ through
\eq
C_0=\frac{2\pi a}{m}\, ,\quad\quad C_2=\frac{\pi a^2r_e}{m}\, .
\en
The bound-state wave function obeys the Schr\"odinger equation
\eq\label{eq:Schroedinger}
({\bf p}^2+\kappa_2^2)\Psi({\bf p})=-m \int \frac{d^d{\bf q}}{(2\pi)^d}\,
H_{int}({\bf p},{\bf q})\Psi({\bf q})\, ,
\en
where $d$ is the dimension of space.

The wave function $\Psi({\bf p})$ consists of a long-range tail and the short-range part, 
which can be approximated by a polynomial. Consequently, one may try to solve
the Eq.~(\ref{eq:Schroedinger}) with the following ansatz:
\eq\label{eq:ansatz}
\Psi({\bf p})=\sqrt{8\pi\kappa_2}\,\frac{{\cal A}_2}{{\bf p}^2+\kappa_2^2}\,
+p_0+p_2{\bf p}^2+\cdots\, .
\en
Substituting this ansatz into the Sch\"odinger equation and performing the integrals in
dimensional regularization with minimal subtraction, it is easy to see that $p_2=p_4=\ldots=0$, while
${\cal A}_2$ and $p_0$ obey the system of linear equations
\eq\label{eq:linear}
\sqrt{8\pi\kappa_2}{\cal A}_2+\kappa_2^2p_0&=&\sqrt{8\pi\kappa_2}
{\cal A}_2\frac{\kappa_2}{4\pi}\biggl(2mC_0-{mC_2}\,\kappa_2^2\biggr)\,,
\nonumber\\[2mm]
p_0&=&{mC_2}\,\sqrt{8\pi\kappa_2}{\cal A}_2\frac{\kappa_2}{4\pi}\, .
\en
Substituting the second equation into the first one, an equation to determine the
bound state momentum $\kappa_2$ emerges
\eq
1-\frac{\kappa_2 m}{2\pi}\,(C_0-C_2\kappa_2^2)=0\, .
\en
The system of the homogeneous linear equations, Eq.~(\ref{eq:linear}) determines only
the ratio $p_0/{\cal A}_2$. In order to determine ${\cal A}_2$, one uses the normalization
condition for the wave function
\eq
 \int\frac{d^d{\bf q}}{(2\pi)^d}\,|\Psi({\bf q})|^2
=\int\frac{d^d{\bf q}}{(2\pi)^d}\,\biggl(\sqrt{8\pi\kappa_2}\,\frac{{\cal A}_2}{{\bf p}^2+\kappa_2^2}\,
+p_0\biggr)^2=1\, .
\en
Evaluating the integrals and expressing $p_0$ through ${\cal A}_2$, we finally get
\eq
{\cal A}_2^2=\biggl(1-\frac{mC_2\kappa_2^3}{\pi}\biggr)^{-1}=\frac{1}{1-\kappa_2r_e}\, ,
\en
where at the final stage we have expressed the non-relativistic couplings through the
physical observables.

Below, we briefly summarize the lessons learned:

\begin{itemize}

\item[(i)]
The energy level shift in a finite volume is determined by the asymptotic part of the
bound state wave function, which is parameterized by two constants $\kappa_2$ and
${\cal A}_2$. 
\item[(ii)]
If the higher-order short-range interactions are absent ($r_e=0$ together with all higher-order terms),
then ${\cal A}_2=1$. This condition is equivalent to Weinberg's compositeness
condition~\cite{Weinberg:1965zz}, which distinguishes a hadronic molecule from a tightly bound compound.
\item[(iii)]
The asymptotic normalization coefficient is determined from the normalization of the
whole wave function to unity. For example, ${\cal A}_2\simeq 0$ would mean that the
short-range component prevails over the long-range one, so that the system is 
predominately a tight compound. In accordance to this, the energy level has a very little
dependence on the volume, and {\em vice versa.}

\end{itemize}

In the next section we shall demonstrate, how the above derivation can be adjusted
for the particle-dimer bound state.

\section{Dimer formalism in the infinite volume}
\label{sec:infinite}

\subsection{The Lagrangian}

In order to simplify things as much as possible, we shall consider the case of three
identical non-relativistic bosons in the CM frame. The inclusion of
relativistic kinematics, spin, moving frames, etc., proceeds along the standard path
and will be addressed in our future publications.

In the following, we shall mainly follow the Refs.~\cite{Bedaque:1998kg,Bedaque:1998km}. The most
general effective Lagrangian that describes the two- and three-particle sectors
is given by
\eq\label{eq:L3}
{\cal L}=\psi^\dagger\biggl(i\partial_0+\frac{\nabla^2}{2m}\biggr)\psi
-\frac{C_0}{2}\,(\psi^\dagger\psi)^2-\frac{D_0}{6}\,(\psi^\dagger\psi)^3+\cdots\, ,
\en
where $\psi$ denotes the non-relativistic field operator for a boson with a mass $m$,
and ellipses stand for the terms with derivatives. We further introduce a dummy field
$T$ (called dimer) with the quantum number of two bosons and consider the Lagrangian
\eq\label{eq:LT}
{\cal L}=\psi^\dagger
\biggl(i\partial_0+\frac{\nabla^2}{2m}\biggr)\psi
+\Delta T^\dagger T-\frac{g}{\sqrt{2}}\,(T^\dagger\psi\psi+\mbox{h.c.})
+hT^\dagger T\psi^\dagger\psi+\cdots\, .
\en
Note that the field $T$ is not dynamical -- the corresponding Lagrangian does not
contain the time derivative. Integrating out this field by using the equations
of motion, we arrive at the Lagrangian
\eq
{\cal L}=\psi^\dagger
\biggl(i\partial_0+\frac{\nabla^2}{2m}\biggr)\psi
-\frac{g^2(\psi^\dagger\psi)^2}{2(\Delta+h\psi^\dagger\psi)}+\cdots\, .
\en
Expanding this Lagrangian in the power of fields, one sees that it describes exactly the
same physics as the Lagrangian from Eq.~(\ref{eq:L3}) in the two- and three-particle
sectors, if the couplings are fixed in the following manner:
\eq
C_0=\frac{g^2}{\Delta}\, ,\quad\quad
D_0=-\frac{3g^2h}{\Delta^2}\, .
\en
The following remarks are in order:
\begin{itemize}
\item[(i)]
As is clear from above, the particle-dimer picture is not an approximation, in fact, 
being restricted to the two and three particle sectors, it is mathematically equivalent
to the original treatment without a dimer field. Hence, the treatment of
the finite-volume effects with the use of the particle-dimer approach is as general as
the one based on a three-particle Lagrangian without a dimer field.  
\item[(ii)]
Using the dimer formalism does not imply the neglect of the higher partial waves.
The two-particle Lagrangian containing derivative terms, which describe P-, D-, $\ldots$
wave interactions, can be replaced by a tower of Lagrangians containing dimers with
angular momentum 1, 2, etc..
The truncation of the partial-wave expansion is then equivalent
to including the dimers with angular momentum below some fixed value (the details of the formalism
can be found in Ref.~\cite{forthcoming}).  
\end{itemize}
One additional remark concerns the inclusion of the kinetic energy term for the dimer.
In principle, $T$ is a dummy field, so, instead of Eq.~(\ref{eq:LT}), one could 
consider the Lagrangian with a dynamical dimer field as well
\eq\label{eq:LD}
{\cal L}=\psi^\dagger
\biggl(i\partial_0+\frac{\nabla^2}{2m}\biggr)\psi
+\sigma T^\dagger \biggl(i\partial_0+\frac{\nabla^2}{4m}-\Delta\biggr) T
-\frac{g}{\sqrt{2}}\,(T^\dagger\psi\psi+\mbox{h.c.})
+hT^\dagger T\psi^\dagger\psi+\cdots\, ,
\en
where $\sigma=\pm 1$ is sign that depends on the sign of the effective range.
The variable $T$ can again be integrated out, leading to an
equivalent theory in terms of a field $\psi$ only. 
However, as argued, e.g., in Ref.~\cite{Kaplan:1996nv}, when
a shallowly bound two-particle state is present, the convergence radius of the perturbation expansion in the
theory with a dynamical dimer should be larger because this theory contains the
small scale $\Delta$ explicitly (not hidden in the couplings of the effective theory).

In the following, we shall use the formulation based on the Lagrangian,
Eq.~(\ref{eq:LT}),
neglecting all higher-order terms. The inclusion of the derivative couplings, higher
partial waves, etc. will be discussed in our forthcoming paper.     

\subsection{The bound-state equation and the normalization condition}

As it is well known, the particle-dimer bound state wave function in the theory
described by the Lagrangian, Eq.~(\ref{eq:LT}), obeys the homogeneous 
Faddeev equation
\eq\label{eq:BS}
\Psi({\bf p})=8\pi \int^\Lambda\frac{d^3{\bf q}}{(2\pi)^3}\,
Z({\bf p},{\bf q};E)\tau({\bf q};E)\Psi({\bf q})\, ,
\en
where $\Lambda$ denotes an explicit UV cutoff, and
\eq
Z({\bf p},{\bf q};E)&=&\frac{1}{-mE+{\bf p}^2+{\bf q}^2+{\bf p}{\bf q}}
+\frac{h}{2mg^2}\, ,
\nonumber\\[2mm]
\tau({\bf q};E)&=&\frac{1}{-a^{-1}+\sqrt{\frac{3}{4}\,{\bf q}^2-mE}}\, .
\en
Projecting to the S-wave and defining $h=2mg^2 H(\Lambda)/\Lambda^2$,
$\kappa^2=-mE$, we arrive at the equation
\eq\label{eq:psi}
\Psi(p)=\frac{4}{\pi}\int_0^\Lambda q^2dq
\biggl\{\frac{1}{2pq}\,\ln\frac{p^2+pq+q^2+\kappa^2}{p^2-pq+q^2+\kappa^2}
+\frac{H(\Lambda)}{\Lambda^2}\biggr\}\tau(q;E)\Psi(q)\, ,
\en
where $\Psi(p)$ stands for the S-wave wave function (note that $\tau({\bf q};E)=\tau(q;E)$ depends only on $q=|{\bf q}|$).
It is well known that  the particle-dimer coupling
constant $H(\Lambda)$ should be a log-periodic function of the cutoff parameter
$\Lambda$ for the limit $\Lambda\to\infty$ to exist in this equation~\cite{Bedaque:1998kg,Bedaque:1998km}.

Next, we shall derive the normalization condition for the wave function $\Psi(p)$, which
has a non-trivial form because the kernel of the integral equation depends on the energy
$E$. The derivation follows the standard pattern~(see, e.g.,~\cite{Itzykson:1980rh}). 
Namely, we consider the inhomogeneous equation for the scattering amplitude
\eq\label{eq:BSA}
{\cal M}({\bf p},{\bf k};E)=Z({\bf p},{\bf k};E)
+8\pi\int^\Lambda\frac{d^3{\bf q}}{(2\pi)^3}\,Z({\bf p},{\bf q};E)\tau({\bf q};E)
{\cal M}({\bf q},{\bf k};E)\, .
\en
In a compact notation, we have ${\cal M}=Z+Z(8\pi\tau){\cal M}$. Defining the Green
function as $G=(8\pi\tau)+(8\pi\tau){\cal M}(8\pi\tau)$, we obtain
$G=(8\pi\tau)+(8\pi\tau)ZG$, and $G^{-1}=(8\pi\tau)^{-1}-Z$.  Further,
using the identity $GG^{-1}G=G$ and the behavior of the Green function in the vicinity
of the bound-state pole
\eq\label{eq:pole}
G({\bf p},{\bf k};E)=\frac{8\pi\tau({\bf p};E_n)\Psi_n({\bf p})\Psi_n({\bf k})8\pi\tau({\bf k};E_n)}{E-E_n}\:+\mbox{ terms regular as $E\to E_n$}\, ,
\en
we arrive at the following normalization condition for the wave function:
\eq\label{eq:norm}
1&=&-8\pi\int^\Lambda\frac{d^3{\bf p}}{(2\pi)^3}\,(\Psi_n({\bf p}))^2
\frac{\partial \tau({\bf p};E)}{\partial E}\biggr|_{E=E_n}
\nonumber\\[2mm]
&-&
(8\pi)^2\int^\Lambda\frac{d^3{\bf p}}{(2\pi)^3}\,\frac{d^3{\bf k}}{(2\pi)^3}\,
\Psi_n({\bf p})\tau({\bf p};E)\frac{\partial Z({\bf p},{\bf k};E)}{\partial E}
\tau({\bf k};E)\Psi_n({\bf k})\biggr|_{E=E_n}\, .
\en

\subsection{Minlos-Faddeev solution}

Assuming $\Lambda\to\infty$ and $H(\Lambda)=0$, we obtain the 
Skornyakov-Ter-Martirosian (STM)
equation~\cite{STM}. Unlike the equation~(\ref{eq:BS}),
the STM equation is known not to possess a unique solution~\cite{Danilov}.
Minlos and Faddeev~\cite{Faddeev-Minlos}
 have found an exact solution to the integral equation in the unitary
limit $a\to \infty$:
\eq\label{eq:psiMF}
\Psi_0\biggl(\frac{p}{\kappa}\biggr)=iN_0\frac{\kappa\sin(s_0u)}{p}\, ,\quad\quad 
u=\ln\biggl(\frac{\sqrt{3}}{2}\,\frac{p}{\kappa}+
\sqrt{\frac{3p^2}{4\kappa^2}+1}\biggr)\, ,
\en
where $\kappa=\sqrt{-mE}$ is the three-body bound state momentum and
$s_0\simeq 1.00624$ is a numerical constant, which is a solution
of the transcendental
equation
\eq
s_0\cosh\frac{\pi s_0}{2}=\frac{8}{\sqrt{3}}\,\sinh\frac{\pi s_0}{6}\, .
\en
We hereafter refer to Eq.~(\ref{eq:psiMF}) as to the Minlos-Faddeev (MF) wave function. 
Note that the function in Eq.~(\ref{eq:psiMF}) is a solution for any value of $\kappa$
-- the spectrum is not quantized.
The overall normalization factor in this equation should be determined from the normalization condition. The substitution of Eq.~(\ref{eq:psiMF}) into (\ref{eq:norm}) gives
 \eq
\frac{\pi}{2mN_0^2}=I_0\, ,
\en
where (see Appendix~\ref{app:integrals})
\eq\label{eq:I_0}
I_0&=&\int_0^\infty dx \frac{\sin^2(s_0u)}{\left(\sqrt{\frac{3x^2}{4}+1}\right)^3}
+\frac{8}{\pi}\int_0^\infty \frac{xdx}{\sqrt{\frac{3x^2}{4}+1}}
\int_0^\infty \frac{ydy}{\sqrt{\frac{3y^2}{4}+1}}
\frac{\sin(s_0u)\sin(s_0v)}{(x^2+y^2+1)^2-x^2y^2}
\nonumber\\[2mm]
&=&\frac{1}{\sqrt{3}}\biggl(1-\frac{\pi s_0}{\sinh\pi s_0}\biggr)
+\frac{8\pi}{9\sinh\pi s_0}\biggl(\sinh\frac{2\pi s_0}{3}-2\sinh\frac{\pi s_0}{3}\biggr)\, ,
\en
and
\eq
u=\ln\biggl(\frac{\sqrt{3}}{2}\,x+\sqrt{\frac{3}{4}\,x^2+1}\biggr)\, ,\quad\quad
v=\ln\biggl(\frac{\sqrt{3}}{2}\,y+\sqrt{\frac{3}{4}\,y^2+1}\biggr)\, .
\en

\subsection{Asymptotic normalization coefficient}

As mentioned above, the STM equation does not have unique solutions. First, imposing a
cutoff, one arrives at the discrete three-particle spectrum. In order to ensure that
the limit $\Lambda\to\infty$ exists, one has to introduce a short-range interaction 
parameterized by a constant $H(\Lambda)$, where the dependence on the cutoff
$\Lambda$ is log-periodic. Furthermore, for a fixed $\Lambda$ and $H(\Lambda)$,
the low-energy spectrum is discrete, condensing towards zero. For
two neighbouring
levels whose energy is much smaller than the cutoff $\Lambda$, 
the following relation holds in the unitary limit:
\eq
\frac{\kappa_{n+1}}{\kappa_n}=\exp(-\pi/s_0)\simeq\frac{1}{22.69}\, .
\en 
Consequently, fixing a single energy level for a given $\Lambda$ is equivalent to the fixing of the parameter $H(\Lambda)$.

The Faddeev equation with a finite cutoff and $H(\Lambda)$ can not be solved
analytically. However, the numerical solution is straightforward. In table~\ref{tab:spectrum}, for
illustration, we give several energy eigenvalues for the choice $\Lambda=10^4$
and $H(\Lambda)=0$.
\begin{table}[t]
\begin{center}
\begin{tabular}{|c|c|c|}
\hline\hline
$n$ & $\kappa_n$ & $\kappa_{n+1}/\kappa_n$ \\
\hline
$1$ & $1779.3756$ & $22.93$ \\
$2$ & $77.5971$ & $22.69$ \\
$3$ & $3.4192$ & $22.69$ \\
$4$ & $0.1507$ & $22.69$ \\
$5$ & $0.006639$ & \\
  & $\ldots$ & \\
\hline\hline
\end{tabular}
\caption{Some energy levels obtained Eq.~(\ref{eq:psi}) 
for the choice
$\Lambda=10^4$ and $H(\Lambda)=0$.}
\label{tab:spectrum}
\end{center}
\end{table}

\begin{figure}[t]
\begin{center}
\includegraphics*[width=10.cm]{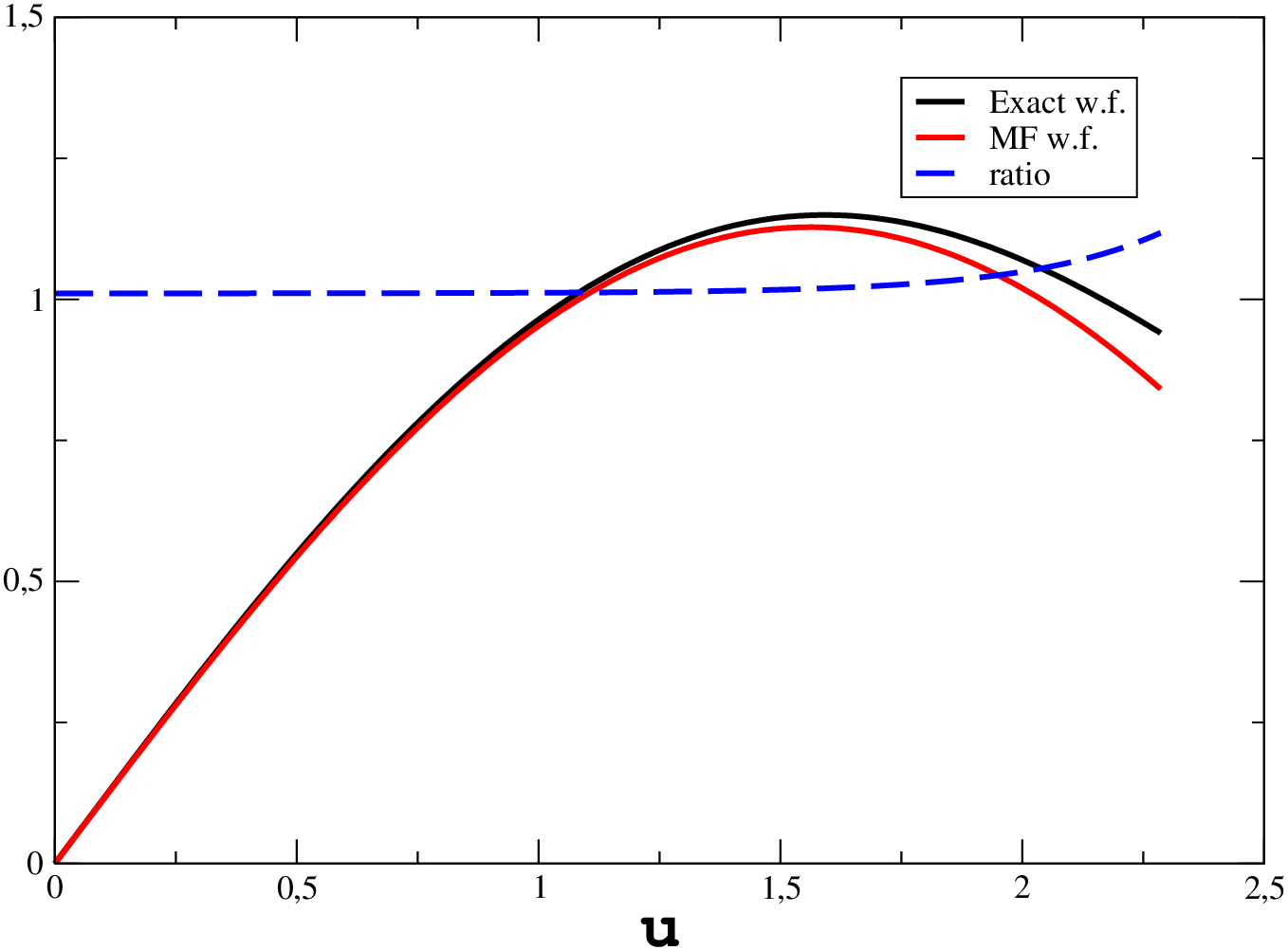}
\end{center}
\caption{The numerical solution of the Faddeev equation vs. the Minlos-Faddeev
wave function. The parameters are taken as $\Lambda=10^4$ and 
$\kappa=1779.3756$, corresponding to $H(\Lambda)=0$. 
We namely 
plot the functions $|p/\kappa\Psi(p/\kappa)|$, $|p/\kappa\Psi_0(p/\kappa)|$
and $A(p/\kappa)$ from Eq.~(\ref{eq:ratio}) vs. the dimensionless
variable $u$, which is defined in Eq.~(\ref{eq:psiMF}). 
The ratio of two functions,
${\cal A}$, is very flat and approaches the value ${\cal A}=1.0105863$ at the origin. }
\label{fig:asymptotic}
\end{figure}

It is clear that, for the momenta much smaller than the cutoff
$\Lambda$, the wave function will be given by the MF solution $\Psi_0$. The difference
can arise only at momenta $p\simeq\Lambda$. The {\em overall normalization,}
however, is a subtler issue since the normalization integral includes all momenta.
To summarize, the solution of the Faddeev equation with cutoff 
{\em at a given bound-state momentum $\kappa$} should be given by
\eq\label{eq:ratio}
\Psi\biggl(\frac{p}{\kappa}\biggr)=A\biggl(\frac{p}{\kappa}\biggr)
\Psi_0\biggl(\frac{p}{\kappa}\biggr)\, ,
\en
where the function $A(x)$ should have a very flat plateau for $x\ll\Lambda/\kappa$.
Then, in analogy to the two-body case, we {\em define} the particle-dimer asymptotic
normalization coefficient as
\eq
{\cal A}=A(0)\, .
\en 
Further, the dimensionless quantity ${\cal A}$ must be a function of the only 
dimensionless combination $\kappa/\Lambda$ that can be composed from the 
parameters of the theory. If $\kappa/\Lambda\to 0$, then, obviously, ${\cal A}\to 1$.
Consequently, the asymptotic normalization coefficient is very close to one for
the shallow three-particle bound states. The explicit demonstration of the above statements
is given in Fig.~\ref{fig:asymptotic}.

The situation changes, when derivative particle-dimer interactions are added. Consider, for example,
adding the term $H_1(\Lambda)({\bf p}^2+{\bf q}^2)$ to the particle-dimer interaction
Hamiltonian (this is an analog of the effective range term in the two-particle case).
The ratio will be still flat for $p\ll \Lambda$. However, the statement
${\cal A}\to 1$, as $\kappa/\Lambda\to 0$ does not hold any more -- in other
words, the asymptotic normalization coefficient encodes the effect of the short-range
physics, as in the two-particle case. In the particle-dimer case, however, 
${\cal A}$ and $H_1(\Lambda)$ can not be related algebraically, and ${\cal A}$ should
be extracted from the numerical solution of the equation with a given value of $H_1(\Lambda)\neq 0$.

\section{Particle-dimer bound state in a finite volume}
\label{sec:finite} 

\subsection{Expression of the first order finite-volume energy shift}

Our derivation -- in the particle-dimer context -- 
will be partly similar to that of Ref.~\cite{Hansen:2016ync}. Consider the 
Faddeev equation in a finite volume
\eq\label{eq:BSAL}
{\cal M}_L({\bf p},{\bf k};E)=Z({\bf p},{\bf k};E)
+\frac{8\pi}{L^3}\sum\limits_{\bf q}\,Z({\bf p},{\bf q};E)\tau_L({\bf q};E)
{\cal M}_L({\bf q},{\bf k};E)\, .
\en
where ${\bf q}=\frac{2\pi}{L}\,{\bf n}\, ,\quad {\bf n}\in \mathbb{Z}^3$ and
\eq\label{eq:tauL}
\tau^{-1}_L({\bf q};E)&=&-a^{-1}+\sqrt{\frac{3}{4}\,{\bf q}^2+\kappa_L^2}
+\Delta_L({\bf q};E)\, ,
\nonumber\\[5mm]
\Delta_L({\bf q};E)&=&-\frac{1}{\pi L}\int d^3{\bf s}\sum\limits_{{\bf n}\neq{\bf 0}}\frac{e^{2\pi i{\bf n} {\bf s}
-i\pi{\bf \hat q}{\bf n}}}{\hat \kappa_L^2+\frac{3}{4}\,{\bf \hat q}^2+{\bf s}^2}\, ,
\quad\quad
{\bf \hat q}=\frac{L{\bf q}}{2\pi}\, ,\quad
\hat \kappa_L=\frac{L\kappa_L}{2\pi}\, .
\en
Here, $\kappa_L^2=-mE$. Moreover, we have assumed that we are below the particle-dimer
breakup threshold, where all denominators are non-singular, and have used
Poisson's summation formula. Note that exactly this equation was considered earlier
in Refs.~\cite{Kreuzer:2010ti,Kreuzer:2009jp,Kreuzer:2008bi,Kreuzer:2012sr}.

The finite-volume effects in the particle-dimer bound state equation emerge at two 
different places. First, the integration over ${\bf q}$ is changed to a 
sum over discrete values in Eq.~(\ref{eq:BSAL}). Second, there is an additional term
$\Delta_L({\bf q};E)$ in Eq.~(\ref{eq:tauL}). Using again Poisson's summation
formula, we may rewrite the Eq.~(\ref{eq:BSAL}) as
\eq\label{eq:BSAP}
{\cal M}_L({\bf p},{\bf k};E)&=&Z({\bf p},{\bf k};E)
+8\pi\int^\Lambda\frac{d^3{\bf q}}{(2\pi)^3}\,Z({\bf p},{\bf q};E)
\hat\tau_L({\bf q};E)
{\cal M}_L({\bf q},{\bf k};E)\, ,
\nonumber\\[2mm]
\hat\tau_L({\bf q};E)&=&
\frac{1+\sum\limits_{{\bf n}\neq{\bf 0}}e^{2i\pi{\bf n}{\bf \hat q}}}
{\tau^{-1}({\bf q};E)+\Delta_L({\bf q};E)}\, .
\en
Excluding now the quantity $Z$ using Eq.~(\ref{eq:BSA}), we obtain
\eq\label{eq:BSAD}
{\cal M}_L({\bf p},{\bf k};E)&=&{\cal M}({\bf p},{\bf k};E)
+8\pi\int^\Lambda\frac{d^3{\bf q}}{(2\pi)^3}\,{\cal M}({\bf p},{\bf k};E)
\delta\tau_L({\bf q};E)
{\cal M}_L({\bf q},{\bf k};E)\, ,
\en
where
\eq
\delta\tau_L({\bf q};E)&=&\hat\tau_L({\bf q};E)-\tau({\bf q};E)
=\sum\limits_{{\bf n}\neq{\bf 0}}e^{2i\pi{\bf n}{\bf \hat q}}\tau({\bf q};E)
\nonumber\\[2mm]
&-&(1+\sum\limits_{{\bf n}\neq{\bf 0}}e^{2i\pi{\bf n}{\bf \hat q}})(\tau({\bf q};E))^2
\Delta_L({\bf q};E)+\cdots\, .
\en
The infinite-volume amplitude ${\cal M}$ has a pole at the bound-state energy
(cf. with Eq,~(\ref{eq:pole})):
\eq\label{eq:poleM}
{\cal M}({\bf p},{\bf k};E)=\frac{
\Psi({\bf p})\Psi({\bf k})}{E-E_L}\:
+\mbox{ terms regular as $E\to E_L$}\, ,\quad\quad
\en
where $\Psi({\bf p})$ is the infinite-volume wave function.
Substituting this ansatz in Eq.~(\ref{eq:BSAD}), we finally obtain the expression for the first-order energy shift of the three-body bound state
\eq
\Delta E_L=8\pi\int^\Lambda\frac{d^3{\bf p}}{(2\pi)^3}\,(\Psi({\bf p}))^2
\delta\tau_L({\bf p};E)\, .
\en

\subsection{Evaluation of the first-order energy shift}

The energy shift can be written as
\eq\label{eq:DeltaE}
\Delta E_L&=&\Delta E_1+\Delta E_2+\cdots\, ,
\nonumber\\[2mm]
\Delta E_1&=&8\pi\int^\Lambda\frac{d^3{\bf p}}{(2\pi)^3}\,
\frac{(\Psi({\bf p}))^2\sum\limits_{{\bf n}\neq{\bf 0}}e^{2i\pi{\bf n}{\bf \hat p}}}
{-a^{-1}+\sqrt{\frac{3}{4}\,{\bf p}^2+\kappa^2}}\, ,
\nonumber\\[2mm]
\Delta E_2&=&-8\pi\int^\Lambda\frac{d^3{\bf p}}{(2\pi)^3}\,
\frac{(\Psi({\bf p}))^2(1+\sum\limits_{{\bf n}\neq{\bf 0}}e^{2i\pi{\bf n}{\bf \hat p}})}
{(-a^{-1}+\sqrt{\frac{3}{4}\,{\bf p}^2+\kappa^2})^2}\,\Delta_L({\bf p},E)\, .
\en
The evaluation of these integrals in the unitary limit proceeds mainly 
along the lines described in Ref.~\cite{Hansen:2016ync}. In this limit, one has $\Lambda\to\infty,~a\to\infty$ and $\Psi({\bf p})$ is the MF wave function $\Psi_0({\bf p})$.
Using Eq.~(\ref{eq:psiMF}) and performing angular integration, we get
\eq
\Delta E_1=-\frac{2N_0^2\kappa^2}{\pi}\int_0^\infty dp\sin^2\biggl(s_0\ln\biggl(\frac{\sqrt{3}}{2}\,\frac{p}{\kappa}
+\sqrt{\frac{3p^2}{4\kappa^2}+1}\biggr)\biggr)
\sum\limits_{{\bf n}\neq{\bf 0}}\frac{e^{iLnp}-e^{-iLnp}}{iLnp}\frac{1}{\sqrt{\frac{3p^2}{4}+\kappa^2}}\, ,
\en
where $n=|{\bf n}|$. It is clear that the leading exponential contribution emerges
from the term with $n=1$. Introducing the variable $u$ defined in Eq.~(\ref{eq:psiMF}),
one gets
\eq
\frac{m\Delta E_1}{\kappa^2}=-\frac{6}{\kappa LI_0}\,
\int_0^\infty \frac{du}{\sinh u}(1-\cos 2s_0u)\sin\biggl(\frac{2\kappa L}{\sqrt{3}}\,\sinh u\biggr)\, .
\en
in $\kappa L\gg 1$, the integral in the r.h.s. of the above equation has the following asymptotic expansion
\eq
\int_0^\infty \frac{du}{\sinh u}(1-\cos 2s_0u)\sin\biggl(\frac{2\kappa L}{\sqrt{3}}\,\sinh u\biggr)=-\frac{ 3^{1/4}\pi^{1/2}}{2\sqrt{\kappa L}}\,(1-\cosh\pi s_0)\exp\biggl(-\frac{2\kappa L}{\sqrt{3}}\biggr)+\cdots\, .
\en
Using this expansion, one reproduces the result first derived in Ref.~\cite{Meissner:2014dea} and re-derived in Ref.~\cite{Hansen:2016ync} (note that in Ref.~\cite{Hansen:2016ync}, an algebraic error contained in the original derivation was corrected):
\eq\label{eq:MRR}
\frac{\Delta E_1}{|E|}=c
(\kappa L)^{-3/2}\exp\biggl(-\frac{2\kappa L}{\sqrt{3}}\biggr)+\cdots\, ,
\en
where
\eq
c=-\frac{2\pi^{1/2}3^{5/4}}{I_0}\sinh^2\frac{\pi s_0}{2}\, .
\en
Taking into account the relation
\eq
I_0=\frac{C_0^{-1}}{6\sqrt{3}\pi^3}\, ,
\en
where $C_0^{-1}$ is defined in Eq.~(16) of Ref.~\cite{Meissner:2014dea}, it is
straightforward to verify that Eq.~(\ref{eq:MRR}) is identical to the final result of
Ref.~\cite{Meissner:2014dea}.\footnote{Note that the $C_0^{-1}$  defined in Eq.~(16) of Ref.~\cite{Meissner:2014dea}
is not related to the $C_0$ defined in Eq.~(\ref{eq:Hint}).}
However, Eq.~(\ref{eq:MRR}) contains more information as the original formula
from Ref.~\cite{Meissner:2014dea}. It corresponds to the unit asymptotic
normalization coefficient ${\cal A}=1$. Now, it is clear, where the non-trivial
three-particle force, encoded in the derivative particle-dimer couplings, will reveal itself:
the $L$-dependence in the formula (\ref{eq:MRR}) remains the same, only the overall
factor will be multiplied by ${\cal A}^2\neq 1$, where ${\cal A}$ can be 
determined from the {\em infinite-volume} solution through the procedure described 
above. The reason for this is that, at small momenta $p\ll\Lambda$, the ratio
$A$ defined in Eq.~(\ref{eq:ratio}) is close to constant and does not affect the large-$L$ behavior of the energy level.

Further, as shown in Ref.~\cite{Hansen:2016ync}, the correction $\Delta E_2$ is subleading and behaves as 
\eq
\Delta E_2\propto (\kappa L)^{-5/2}\exp\biggl(-\frac{2\kappa L}{\sqrt{3}}\biggr)
\en
for a large $L$. 
The subsequent terms are even more suppressed.

To summarize, we have reproduced the result of 
Ref.~\cite{Meissner:2014dea} for the leading finite-volume energy shift of the 
three-body bound state in the unitary limit  in the particle-dimer picture. Moreover, we have shown that in the 
unitary limit, the asymptotic normalization coefficient emerges from three-particle
derivative forces, and this coefficient is equal to one if such forces are absent.

At the next step, we shall investigate the system beyond the unitary limit.

\section{Beyond the unitary limit}
\label{sec:beyond}

\subsection{Energy shift}

As seen,  the correction $\Delta E_1$ given by Eq.~(\ref{eq:DeltaE}), gives the leading
contribution to the finite-volume energy level in the unitary limit. We expect that this
statement stays valid for finite values of $a$. Singling out the contribution with $|{\bf n}|=1$, one may rewrite the leading contribution to $\Delta E$ in the following form
\eq
 \Delta E\propto \int^\Lambda\frac{d^3{\bf p}}{(2\pi)^3}\,
\frac{(\Psi({\bf p}))^2e^{2i\pi{\bf n}{\bf \hat p}}}
{-a^{-1}+\sqrt{\frac{3}{4}\,{\bf p}^2+\kappa^2}}+\cdots\, ,\quad\quad |{\bf n}|=1\, .
\en
As we shall demonstrate below, the wave function $\Psi({\bf p})$ is regular near origin
(more precisely, the singularities of $\Psi({\bf p})$ are located much farther from the 
origin than the singularities of the denominator). This means that the singularities
of $\Psi({\bf p})$ do not contribute to the large-$L$ behavior of the energy shift
at leading order and hence, at this order, $\Psi({\bf p})$ can be replaced by a constant.
Performing the angular integration, we arrive at the following result
\eq\label{eq:beyond}
\Delta E\propto \frac{1}{L}\,\int_{-\infty}^\infty\frac{pdp}{2\pi i}\,
\frac{e^{ipL}(a^{-1}+\sqrt{\frac{3}{4}\,{\bf p}^2+\kappa^2})}
{\frac{3}{4}\,{\bf p}^2+\kappa^2-a^{-2}}\, .
\en
Note that the quantity $\kappa^2-a^{-2}$ is always positive, if a bound state of 
a particle and a bound dimer is considered (recall that $\kappa_2=a^{-1}$ is the binding
momentum of the dimer in the unitary limit). One has to distinguish two limiting cases:

\subsubsection*{A shallow bound state of a particle and a deeply bound dimer}
In this case, we have $\kappa^2-a^{-2}\ll\kappa^2$. The singularity
at $p=\pm i\sqrt{\frac{4}{3}\,(\kappa^2-a^{-2})}$ is dominant, and the singularity
arising from the square root (cut) can be neglected. Performing the Cauchy integration,
we get
\eq\label{eq:L-1}
\Delta E\propto\frac{1}{L}\, \exp\biggl(-\frac{2}{\sqrt{3}}\,\sqrt{\kappa^2-a^{-2}}L\biggr)\, .
\en
In other words, we reproduce L\"uscher's original result for a two particle (particle-dimer)
bound state~\cite{Luescher-1}.
Note also that this is in a complete agreement with the result of the recent paper~\cite{Konig:2017krd}.

\subsubsection*{A shallow bound state of three particles}

This corresponds to the opposite limit $\kappa^2\gg a^{-2}$. Then, the first term
in Eq.~(\ref{eq:beyond}) is very small and the energy shift is dominated by the 
second term. It is straightforward to see that, in this case,
\eq\label{eq:L-32}
\Delta E\propto\frac{1}{L^{3/2}}\, \exp\biggl(-\frac{2}{\sqrt{3}}\,\kappa L\biggr)\, .
\en
In other words, the result of Ref.~\cite{Meissner:2014dea} is reproduced in this limit.
To be more precise, for any small but finite value of $a^{-2}$
the asymptotic behavior of the energy shift is still given by Eq.~(\ref{eq:L-1}).
However, the coefficient of the leading term is very small, whereas the coefficient in front of
the subleading term given by Eq.~(\ref{eq:L-32}) is of order of unity. So, for large
(but not asymptotically large) values of $L$ the behavior is given by Eq.~(\ref{eq:L-32}),
whereas  Eq.~(\ref{eq:L-1}) sets in asymptotically.

\subsection{Wave function}

The wave function obeys the equation~(\ref{eq:psi}). The location of the singularities
of $\Psi(p)$ in the complex-$p$ plane is determined, as usual, by the Landau equations.
There are two types of singularities (note that the denominator
$-a^{-1}+\sqrt{\frac{3q^2}{4}+\kappa^2}$ does not vanish in the integration
region):
\subsubsection*{Endpoint singularities}

\begin{figure}[t]
\begin{center}
\includegraphics*[width=8.cm]{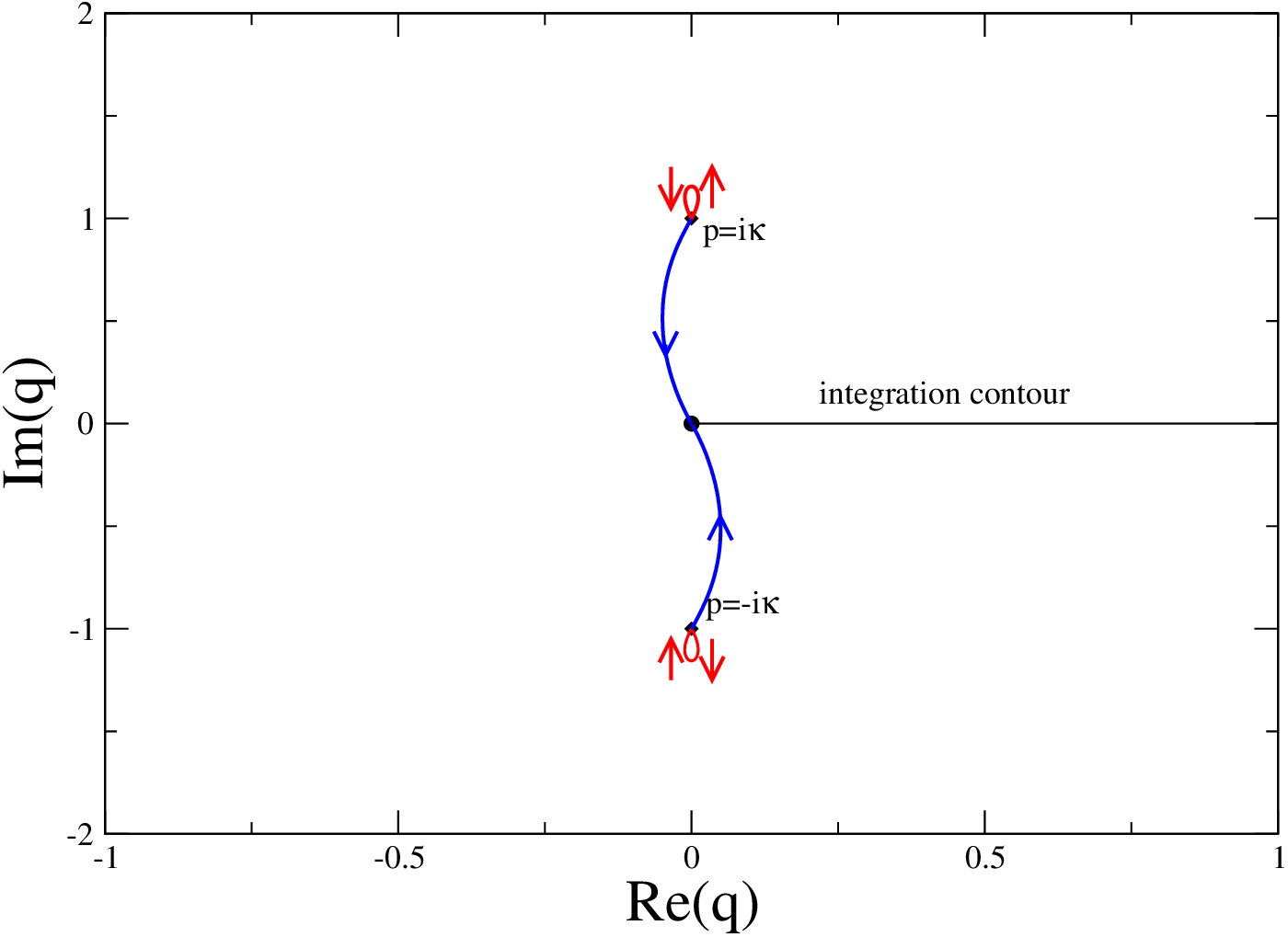}
\vspace*{.8cm}
\includegraphics*[width=8.cm]{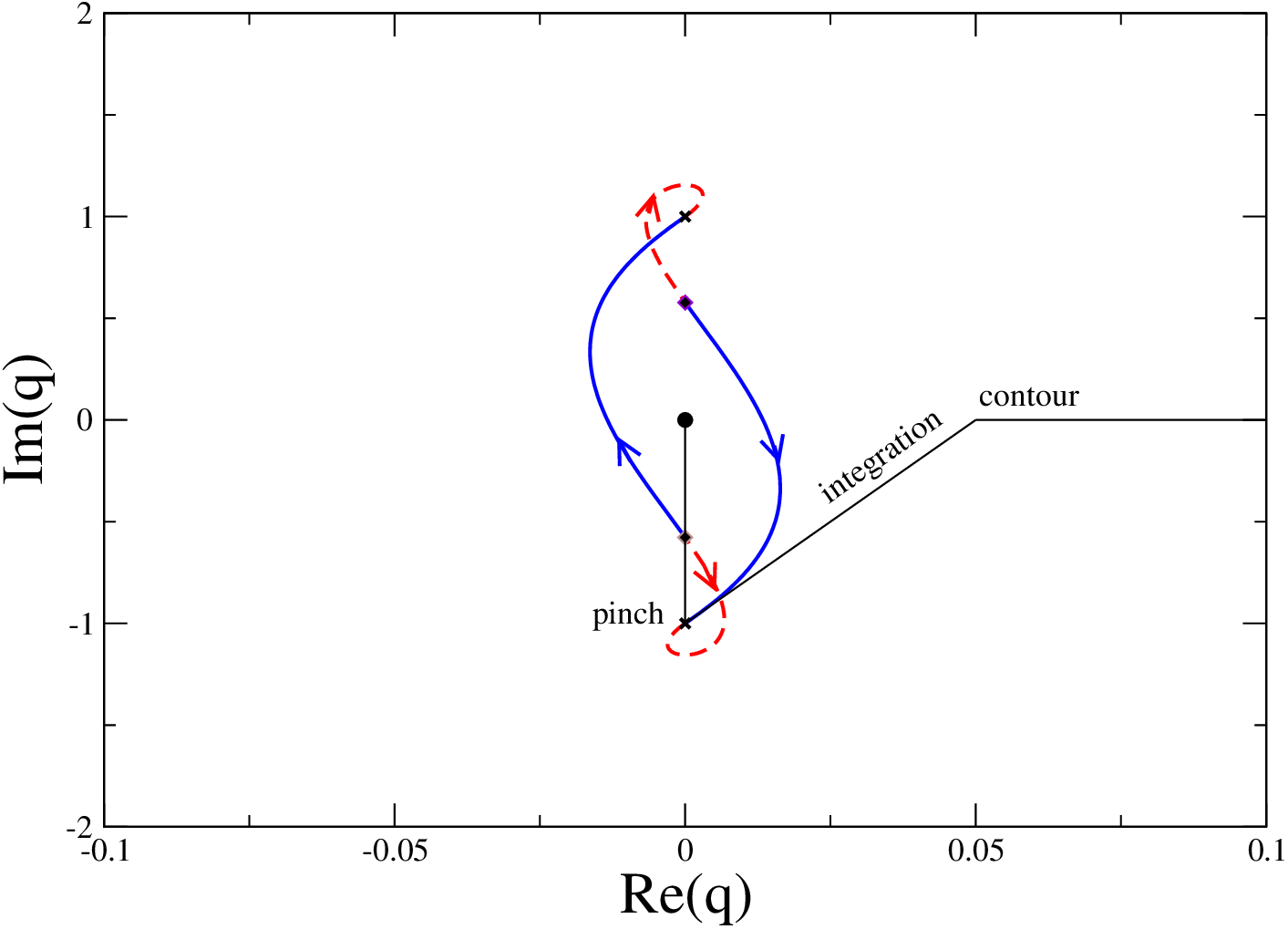}
\end{center}
\caption{Trajectories of the singularities of the kernel of Eq.~(\ref{eq:psi}) in the complex $q$-plane,
Left panel: $p\to i\kappa$, right panel: $p\to \frac{2i}{\sqrt{3}}\,\kappa$.}
\label{fig:trajectories}
\end{figure}

The argument of the logarithm is
$a_\pm(q)=p^2\pm p\cdot q+q^2+\kappa^2$. At $q=0$, the equation
$a_\pm(q=0)=0$ yields $p=\pm i\kappa$. Examine now this potential singularity 
in detail. Let us start, for instance, at $p=0$ and approach the singular point
$p\to i\kappa$ along some path in the complex $p$-plane (for instance,
along the path $p=it+0.05t(1-t),~0\leq t\leq 1$). The four singularities of the 
logarithm, which are determined by the solutions of the equations $a_\pm(q)=0$,
 travel along the lines
\eq
q_{1,2}(p)=\frac{-p\pm i\sqrt{3p^2+4\kappa^2}}{2}\, ,\quad\quad
q_{3,4}(p)=\frac{p\pm i\sqrt{3p^2+4\kappa^2}}{2}\, ,
\en
whereas the singularity of the denominator given by the 
equation $-a^{-1}+\sqrt{\frac{3}{4}\,q^2+\kappa^2}=0$ stays fixed in the 
in the complex $q$-plane. The trajectories $q_{1,2,3,4}$ are shown schematically
in Fig.~\ref{fig:trajectories}, left panel. Two singularities travel from $q=\pm i\kappa$
towards $q=0$ and two others return to $q=\pm i\kappa$. The contour deformation is not needed. 
Substituting now $p=\pm i\kappa$ into the kernel,
we get
\eq
\pm\frac{1}{2i\kappa q}\,\ln\frac{\pm i\kappa+q}{\mp i\kappa+q}
=\frac{1}{\kappa q}\arctan\frac{\kappa}{q}\, .
\en
The logarithm is indeed singular at $q=0$, but 
the integral over $q$ exists, due to the presence of an additional factor $q^2$. 
Consequently, there is no singularity at $p=\pm i\kappa$.

Further, at $q=\Lambda$,
 we get the equation $p^2\pm p\Lambda+\Lambda^2+\kappa^2=0$.
The solution of this equation gives $p=\pm\frac{1}{2}\,(\Lambda\pm
\sqrt{\Lambda^2-4(\Lambda^2+\kappa^2)})$. These points are located very far
from the origin and should not be taken into account.

\subsubsection*{Pinch singularities}

In order to find the location of the pinch singularities, we have to solve the equation
\eq
\frac{d}{dq}\,(p^2\pm pq+q^2+\kappa^2)=0\, .
\en
this gives $p=\pm 2q$. Substituting this back to the argument of the logarithm, 
we get $a_\pm(q)=\left( q\pm \frac{p}{2}\right)^2+\frac{3}{4}\,p^2+\kappa^2$,
i.e., the argument vanishes at $p=\pm \frac{2i}{\sqrt{3}}\,\kappa$.

Consider again the trajectories of the singularities of the logarithm in the complex
$q$-plane, when $p$ varies, according to, for instance, along the path 
 $p=2/\sqrt{3}(it+0.05t(1-t)),~0\leq t\leq 1$.
These trajectories are shown in Fig.~\ref{fig:trajectories}, right panel. In this case,
it is necessary to deform the integration contour, in order to avoid the singularities of the
logarithm. At $p=\pm \frac{2i}{\sqrt{3}}\kappa$ the contour gets pinched between two 
singularities. It is, however, straightforward to check that the singularity of the integrand
along the new integration contour is of an integrable type: logarithmic for 
$a^{-1}\neq 0$ and of square-root type for $a^{-1}=0$. Consequently, the function
$\Psi(p)$, defined by this integral, is non-singular there (albeit the derivatives become,
in general, singular).

\section{Conclusions}
\label{sec:concl}

Our conclusions are as follows:

\begin{itemize}

\item[(i)]
In this paper,  we have rederived the well-known result~\cite{Meissner:2014dea}
for the leading-order finite volume 
energy shift of a shallow three-particle bound state in the unitary limit
using the dimer formalism. While this result
was not unexpected, since the particle-dimer picture is {\em algebraically equivalent}
to the three-particle description, it provides a useful check on the particle-dimer
formalism in a finite volume.

\item[(ii)]
Our treatment goes beyond Refs.~\cite{Meissner:2014dea,Hansen:2016ync}.
Namely, we concentrate on the role of the three-particle force, which is necessary
to carry out the renormalization program in the infinite volume. We have shown that
the algebraic form of the leading-order formula does not change in the presence of
the three-particle force, and only the numerical value of the particle-dimer asymptotic
normalization constant is altered. This constant is equal to one for the STM equation
and differs from unity in the presence of the derivative three-particle interactions --
similar to the two-body case.

\item[(iii)]
Finally, we go beyond the unitary limit and derive the leading-order
formula in this case. This formula smoothly interpolates between two extremes:
the well-known three-particle bound state in the unitary limit
and the bound state of a particle and a deeply bound dimer, for which the usual
L\"uscher formula applies. The study of these limits enables us to
explore the region of applicability of the energy shift formula from 
Ref.~\cite{Meissner:2014dea}.

\item[(iv)]
A host of additional effects awaits to be included, namely, the effective range 
expansion in the two particle sector, higher partial waves and partial wave mixing, 
non-rest frames, relativistic kinematics, etc. Moreover, a general and {\em tractable} quantization
condition, which could be used by lattice practitioners to
analyze the data in the three-particle sector, remains to be worked out. The particle-dimer
language allow one to achieve most of the above goals with an impressive ease.
However, we relegate the proof of this statement to our forthcoming 
publication~\cite{forthcoming}.

\end{itemize}

\begin{acknowledgments}
The authors would like to thank R. Briceno, Z. Davoudi, M. D\"oring, 
E. Epelbaum, M. Hansen, D. Lee, T. Luu, M. Mai, U.-G. Mei{\ss}ner and S. Sharpe 
for useful discussions and D. Lee and U.-G. Mei{\ss}ner for comments on
the manuscript.  

We acknowledge support from the DFG through funds provided to the
Sino-German CRC 110 ``Symmetries and the Emergence of Structure in QCD'' and the
CRC 1245 ``Nuclei: From Fundamental Interactions to Structure and Stars'' as well as
the BMBF under contract 05P15RDFN1.
This research is also supported in part by Volkswagenstiftung
under contract no. 86260 
and by Shota Rustaveli National Science Foundation (SRNSF), grant no. DI-2016-26.

\end{acknowledgments}

\appendix

\section{Two-body problem using power divergence subtraction}
\label{app:cutoff}

In Sec.~\ref{sec:twobody}, we have discussed the energy level shift of a shallow 
two-body bound state and showed that the whole effect of the short-range interactions is
concentrated in the two-body asymptotic normalization coefficient ${\cal A}_2$.
Dimensional regularization with minimal subtraction was used in the derivation.

In this appendix, we 
demonstrate that the same result is obtained in dimensional regularization 
with power divergence subtraction~\cite{Kaplan:1998tg}, where poles in $1/(d-3)$ and in $1/(d-2)$
are subtracted from the integrals ($d$ is the number of spatial dimensions). This generates a non-trivial dependence
of the couplings $C_0$ and $C_2$ on the renormalization scale $\mu$
which must cancel in physical observables.\footnote{We have also checked that the same result 
  is obtained when divergent intergrals are regularized with a momentum cutoff $\Lambda$ but we refrain
  from showing explicit expressions here. In this case,
  divergences up to fifth order in the cutoff $\Lambda$ appear. Moreover, lower-order
  couplings are renormalized by higher orders, which leads to more complicated equations.}

The Schr\"odinger equation for the wave function in the S-wave is rewritten as
\eq
(p^2+\kappa_2^2)\Psi(p)=-m\int\frac{d^d{\bf q}}{(2\pi)^d}\,H_{int}(p,q)\Psi(q)\, ,
\quad H(p,q)=2C_0(\mu)+C_2(\mu)\,(p^2+q^2)+\ldots\, .
\en
Here, $\mu$ denotes the renormalization scale. The coupling constants $C_0(\mu)$ and
$C_2(\mu)$ can be determined from matching to the effective range expansion for the
two-body scattering amplitude. We obtain
\eq
C_0(\mu)=\frac{2\pi}{m}\left(\frac{1}{a}-\mu\right)^{-1}\, ,\quad\quad C_2(\mu)=\frac{m}{2\pi}\,C_0(\mu)^2\,\frac{r_e}{2}\, ,
\label{eq:runPDS}
\en
where $|a|\gg r_e$ was assumed.

Using again the ansatz from Eq.~(\ref{eq:ansatz}) with $p_2=\ldots=0$, we get
\eq
p_0=\frac{mC_2(\mu)}{4\pi}\,\sqrt{8\pi\kappa_2}\,{\cal A}_2\,\left[\kappa_2-\mu\right]\, .
\en
The equation for the bound-state momentum $\kappa_2$ takes the form
\eq
1
&=&\frac{mC_0(\mu)}{2\pi}\,\left[\kappa_2 -\mu\right] -\frac{mC_2(\mu)}{2\pi}\,\kappa_2^2\,\left[\kappa_2 -\mu\right]\,.
\en
Inserting Eqs.~(\ref{eq:runPDS}), we can rewrite this expression as
\eq
\kappa_2=\frac{1}{a}+\frac{r_e}{2}\kappa_2^2+O(\kappa_2^4)\,.
\en

The normalization condition,
\eq
\int \frac{d^d{\bf q}}{(2\pi)^d}\,\biggl(\frac{\sqrt{8\pi\kappa_2}{\cal A}_2}
{q^2+\kappa_2^2}+p_0\biggr)^2=1\,,
\en
yields the following expression for the asymptotic normalization coefficient:
\eq
   {\cal A}_2^{-2}=1-\kappa_2 r_e \frac{\left[\kappa_2 -\mu\right]^2}{\left[1/a -\mu\right]^2}\,.
\en
Using the equation $a\kappa_2=1+O(\kappa_2^3)$, one may rewrite the above equation as
\eq
{\cal A}_2^{-2}=1-\kappa_2 r_e+O(\kappa_2^3)\, .
\en
As we see, the final result for ${\cal A}_2$ does not depend on the regularization used.

\section{Calculation of the integrals}
\label{app:integrals}

In order to calculate the integrals in Eq.~(\ref{eq:I_0}), it is convenient to change
the integration variables
\eq
\frac{\sqrt{3}}{2}\,x=\sinh u\, ,\quad\quad
\frac{\sqrt{3}}{2}\,y=\sinh v\, .
\en
Then, $I_0=I_1+I_2$, where
\eq
I_1=\frac{2}{\sqrt{3}}\,\int_0^\infty du\frac{\sin^2(s_0u)}{\cosh^2u}
=\frac{1}{\sqrt{3}}\,\biggl(1-\frac{\pi s_0}{\sinh(\pi s_0)}\biggr)\, ,
\en
where the last equality was obtained by using the equality given in Ref.~\cite{Gradstein}
\eq
\int_0^\infty \sin ax\frac{\sinh\beta x}{\cosh^2\gamma x}dx=
\frac{\pi\biggl(a\sin\frac{\beta\pi}{2\gamma}\cosh\frac{a\pi}{2\gamma}
-\beta\cos\frac{\beta\pi}{2\gamma}\sinh\frac{a\pi}{2\gamma}\biggr)}
{\gamma^2\biggl(\cosh\frac{a\pi}{\gamma}-\cos\frac{\beta\pi}{\gamma}\biggr)}
\en
with $a=s_0,~\beta=is_0,~\gamma=1$.

Using the same substitution, we obtain
\eq
I_2=\frac{2}{\pi}\,\int_{-\infty}^\infty \int_{-\infty}^\infty dudv
\frac{\sin(s_0u)\sin(s_0v)}{\sinh^2u+\sinh^2v-\sinh u\sinh v+\frac{3}{4}}\, .
\en
It is convenient to define $w=u+v$ and $z=u-v$. Then,
\eq
I_2=\frac{2}{\pi}\,\int_{-\infty}^\infty dz \cos(2s_0z)J(z)\, ,
\en
where
\eq\label{eq:Jz}
J(z)=\frac{4\sqrt{3}\pi}{9}\,\biggl(\frac{1}{\cosh^2z+3\sinh^2z}
-\frac{2}{3\cosh^2z+\sinh^2z}\biggr)\, .
\en
The integral over the variable $z$ can again be performed, using the formula from 
Ref.~\cite{Gradstein}
\eq
\int_0^\infty \frac{\cos ax dx}{\cosh\beta x+\cos\gamma}=\frac{\pi}{\beta}\,
\frac{\sinh\frac{a\gamma}{\beta}}{\sin\gamma\sinh\frac{a\pi}{\beta}}
\en
with $a=2s_0,~\beta=2$ and $\gamma=\frac{2\pi}{3}$ or $\gamma=\frac{\pi}{3}$
(in the first and the second terms of Eq.~(\ref{eq:Jz}), respectively).
At the end, one gets
\eq
I_2=\frac{8\pi}{9}\,\frac{1}{\sinh(\pi s_0)}\,
\biggl(\sinh\frac{2\pi s_0}{3}-2\sinh\frac{\pi s_0}{3}\biggr)\, ,
\en
and the Eq.~(\ref{eq:I_0}) is reproduced.


\begin{thebibliography}{100}

\bibitem{Mathur:2003zf}
  N.~Mathur, Y.~Chen, S.~J.~Dong, T.~Draper, I.~Horvath, F.~X.~Lee, K.~F.~Liu and J.~B.~Zhang,
  Phys.\ Lett.\ B {\bf 605} (2005) 137,
  \href{http://arXiv.org/abs/hep-ph/0306199}{arXiv:hep-ph/0306199}.

\bibitem{Guadagnoli:2004wm}
  D.~Guadagnoli, M.~Papinutto and S.~Simula,
  Phys.\ Lett.\ B {\bf 604} (2004) 74,\\
  \href{http://arXiv.org/abs/hep-lat/0409011}{arXiv:hep-lat/0409011}.
  
\bibitem{Leinweber:2004it}
  D.~B.~Leinweber, W.~Melnitchouk, D.~G.~Richards, A.~G.~Williams and J.~M.~Zanotti,
  Lect.\ Notes Phys.\  {\bf 663} (2005) 71,
  \href{http://arXiv.org/abs/nucl-th/0406032}{arXiv:nucl-th/0406032}.

\bibitem{Sasaki:2005ap}
  K.~Sasaki, S.~Sasaki and T.~Hatsuda,
  Phys.\ Lett.\ B {\bf 623} (2005) 208,
  \href{http://arXiv.org/abs/hep-lat/0504020}{arXiv:hep-lat/0504020}.
\bibitem{Sasaki:2005ug}
  K.~Sasaki and S.~Sasaki,
  Phys.\ Rev.\ D {\bf 72} (2005) 034502,
  \href{http://arXiv.org/abs/hep-lat/0503026}{arXiv:hep-lat/0503026}.

\bibitem{Burch:2006cc}
  T.~Burch, C.~Gattringer, L.~Y.~Glozman, C.~Hagen, D.~Hierl, C.~B.~Lang and A.~Schafer,
  Phys.\ Rev.\ D {\bf 74} (2006) 014504,
  \href{http://arXiv.org/abs/hep-lat/0604019}{arXiv:hep-lat/0604019}.


\bibitem{Liu:2014jua}
  K.~F.~Liu, Y.~Chen, M.~Gong, R.~Sufian, M.~Sun and A.~Li,
  PoS LATTICE {\bf 2013} (2014) 507,\\
  \href{http://arXiv.org/abs/1403.6847}{arXiv:1403.6847 [hep-ph]}.

\bibitem{Mahbub:2010jz}
  M.~S.~Mahbub, A.~O.~Cais, W.~Kamleh, D.~B.~Leinweber and A.~G.~Williams,
  Phys.\ Rev.\ D {\bf 82} (2010) 094504,
  \href{http://arXiv.org/abs/1004.5455}{arXiv:1004.5455 [hep-lat]}.

\bibitem{Lasscock:2007ce}
  B.~G.~Lasscock, J.~N.~Hedditch, W.~Kamleh, D.~B.~Leinweber, W.~Melnitchouk, A.~G.~Williams and J.~M.~Zanotti,
  Phys.\ Rev.\ D {\bf 76} (2007) 054510,
  \href{http://arXiv.org/abs/0705.0861}{arXiv:0705.0861 [hep-lat]}.


\bibitem{Lang:2016hnn}
  C.~B.~Lang, L.~Leskovec, M.~Padmanath and S.~Prelovsek,
  Phys.\ Rev.\ D {\bf 95} (2017) 014510,\\
  \href{http://arXiv.org/abs/1610.01422}{arXiv:1610.01422 [hep-lat]}.


\bibitem{Beane:2010em}
  S.~R.~Beane, W.~Detmold, K.~Orginos and M.~J.~Savage,
  Prog.\ Part.\ Nucl.\ Phys.\  {\bf 66} (2011) 1,\\
  \href{http://arXiv.org/abs/1004.2935}{arXiv:1004.2935 [hep-lat]}.

\bibitem{Beane:2012vq}
  S.~R.~Beane {\it et al.} [NPLQCD Collaboration],
  Phys.\ Rev.\ D {\bf 87} (2013) 034506,\\
  \href{http://arXiv.org/abs/1206.5219}{arXiv:1206.5219 [hep-lat]}.

\bibitem{Chang:2015qxa}
  E.~Chang {\it et al.} [NPLQCD Collaboration],
  Phys.\ Rev.\ D {\bf 92} (2015) 114502,\\
  \href{http://arXiv.org/abs/1506.05518}{arXiv:1506.05518 [hep-lat]}.


\bibitem{Epelbaum:2009pd}
  E.~Epelbaum, H.~Krebs, D.~Lee and U.-G.~Mei{\ss}ner,
  Phys.\ Rev.\ Lett.\  {\bf 104} (2010) 142501,\\
  \href{http://arXiv.org/abs/0912.4195}{arXiv:0912.4195 [nucl-th]}.

\bibitem{Epelbaum:2011md}
  E.~Epelbaum, H.~Krebs, D.~Lee and U.-G.~Mei\ss ner,
  Phys.\ Rev.\ Lett.\  {\bf 106} (2011) 192501,\\
  \href{http://arXiv.org/abs/1101.2547}{arXiv:1101.2547 [nucl-th]}.

  
\bibitem{Rokash:2013xda}
  A.~Rokash, E.~Epelbaum, H.~Krebs, D.~Lee and U.-G.~Mei{\ss}ner,
  J.\ Phys.\ G {\bf 41} (2014) 015105,\\
  \href{http://arXiv.org/abs/1308.3386}{arXiv:1308.3386 [nucl-th]}.

\bibitem{Elhatisari:2015iga}
  S.~Elhatisari, D.~Lee, G.~Rupak, E.~Epelbaum, H.~Krebs, T.~A.~L\"ahde, T.~Luu and U.-G.~Mei{\ss}ner,
  Nature {\bf 528} (2015) 111,
  \href{http://arXiv.org/abs/1506.03513}{arXiv:1506.03513 [nucl-th]}.

\bibitem{Polejaeva:2012ut}
  K.~Polejaeva and A.~Rusetsky,
  Eur.\ Phys.\ J.\ A {\bf 48} (2012) 67,
  \href{http://arXiv.org/abs/1203.1241}{arXiv:1203.1241 [hep-lat]}.

\bibitem{Meissner:2014dea}
  U.-G.~Mei{\ss}ner, G.-Rios and A.~Rusetsky,
  Phys.\ Rev.\ Lett.\  {\bf 114} (2015) 091602,
  [Erratum: Phys.\ Rev.\ Lett.\  {\bf 117} (2016) 069902],
  \href{http://arXiv.org/abs/1412.4969}{arXiv:1412.4969 [hep-lat]}.



\bibitem{Guo:2016fgl}
  P.~Guo,
  Phys.\ Rev.\ D {\bf 95} (2017) 054508,
  \href{http://arXiv.org/abs/1607.03184}{arXiv:1607.03184 [hep-lat]}.

\bibitem{Guo:2017ism}
  P.~Guo and V.~Gasparian,
  \href{http://arXiv.org/abs/1701.00438}{arXiv:1701.00438 [hep-lat]}.


\bibitem{Briceno:2012rv}
  R.~A.~Briceno and Z.~Davoudi,
  Phys.\ Rev.\ D {\bf 87} (2013) 094507,
  \href{http://arXiv.org/abs/1212.3398}{arXiv:1212.3398 [hep-lat]}.



\bibitem{Hansen:2014eka}
  M.~T.~Hansen and S.~R.~Sharpe,
  Phys.\ Rev.\ D {\bf 90} (2014) 116003,
  \href{http://arXiv.org/abs/1408.5933}{arXiv:1408.5933 [hep-lat]}.

\bibitem{Hansen:2015zga}
  M.~T.~Hansen and S.~R.~Sharpe,
  Phys.\ Rev.\ D {\bf 92} (2015) 114509,
  \href{http://arXiv.org/abs/1504.04248}{arXiv:1504.04248 [hep-lat]}.

\bibitem{Hansen:2015zta}
  M.~T.~Hansen and S.~R.~Sharpe,
  Phys.\ Rev.\ D {\bf 93} (2016) 014506,
  \href{http://arXiv.org/abs/1509.07929}{arXiv:1509.07929 [hep-lat]}.

\bibitem{Hansen:2016fzj}
  M.~T.~Hansen and S.~R.~Sharpe,
  Phys.\ Rev.\ D {\bf 93} (2016) 096006,
  \href{http://arXiv.org/abs/1602.00324}{arXiv:1602.00324 [hep-lat]}.

\bibitem{Hansen:2016ync}
  M.~T.~Hansen and S.~R.~Sharpe,
  Phys.\ Rev.\ D {\bf 95} (2017) 034501,
  \href{http://arXiv.org/abs/1609.04317}{arXiv:1609.04317 [hep-lat]}.


\bibitem{Briceno:2017tce}
  R.~A.~Brice\~no, M.~T.~Hansen and S.~R.~Sharpe,
  Phys. \ Rev. \ D {\bf 95} (2017) 074510,\\
  \href{http://arXiv.org/abs/1701.07465}{arXiv:1701.07465 [hep-lat]}.

\bibitem{Luescher-torus}
  M.~Luscher,
  Nucl.\ Phys.\ B {\bf 354} (1991) 531,
  \href{https://doi.org/10.1016/0550-3213(91)90366-6}{doi:10.1016/0550-3213(91)90366-6}.

\bibitem{Kreuzer:2010ti}
  S.~Kreuzer and H.-W.~Hammer,
  Phys.\ Lett.\ B {\bf 694} (2011) 424,
  \href{http://arXiv.org/abs/1008.4499}{arXiv:1008.4499 [hep-lat]}.

\bibitem{Kreuzer:2009jp}
  S.~Kreuzer and H.-W.~Hammer,
  Eur.\ Phys.\ J.\ A {\bf 43} (2010) 229,
  \href{http://arXiv.org/abs/0910.2191}{arXiv:0910.2191 [nucl-th]}.

\bibitem{Kreuzer:2008bi}
  S.~Kreuzer and H.-W.~Hammer,
  Phys.\ Lett.\ B {\bf 673} (2009) 260,
  \href{http://arXiv.org/abs/0811.0159}{arXiv:0811.0159 [nucl-th]}.

\bibitem{Kreuzer:2012sr}
  S.~Kreuzer and H.~W.~Grie{\ss}hammer,
  Eur.\ Phys.\ J.\ A {\bf 48} (2012) 93,
  \href{http://arXiv.org/abs/1205.0277}{arXiv:1205.0277 [nucl-th]}.


\bibitem{Bedaque:1998kg}
  P.~F.~Bedaque, H.-W.~Hammer and U.~van Kolck,
  Phys.\ Rev.\ Lett.\  {\bf 82} (1999) 463,\\
  \href{http://arXiv.org/abs/nucl-th/9809025}{arXiv:nucl-th/9809025}.

\bibitem{Bedaque:1998km}
  P.~F.~Bedaque, H.-W.~Hammer and U.~van Kolck,
  Nucl.\ Phys.\ A {\bf 646} (1999) 444,\\
  \href{http://arXiv.org/abs/nucl-th/9811046}{arXiv:nucl-th/9811046}.
\bibitem{Epelbaum:2016ffd}
  E.~Epelbaum, J.~Gegelia, U.-G.~Mei\ss ner and D.~L.~Yao,
  Eur.\ Phys.\ J.\ A {\bf 53} (2017)  98,
  \href{http://arXiv.org/abs/1611.06040}{arXiv:1611.06040 [nucl-th]}.


  
\bibitem{forthcoming}
H.-W. Hammer, J.-Y. Pang and A. Rusetsky, in preparation.

\bibitem{Luescher-1}
  M.~L\"uscher,
  Commun.\ Math.\ Phys.\  {\bf 104} (1986) 177,
  \href{https://doi.org/10.1007/BF01211589}{doi:10.1007/BF01211589}.

\bibitem{Weinberg:1965zz}
  S.~Weinberg,
  Phys.\ Rev.\  {\bf 137} (1965) B672,
  \href{https://doi.org/10.1103/PhysRev.137.B672}{doi:10.1103/PhysRev.137.B672}.

\bibitem{Kaplan:1996nv}
  D.~B.~Kaplan,
  Nucl.\ Phys.\ B {\bf 494} (1997) 471,
  \href{http://arXiv.org/abs/nucl-th/9610052}{arXiv:nucl-th/9610052}.

\bibitem{Itzykson:1980rh}
  C.~Itzykson and J.~B.~Zuber,
  {\em Quantum Field Theory,}
  McGraw-Hill (1980).

\bibitem{STM}
G.~V.~Skornyakov and K.~A.~Ter-Martirosyan, Zh. Eksp. Teor.
Fiz. {\bf 31} (1956) 775,\\
 \href{http://www.jetp.ac.ru/cgi-bin/e/index/e/4/5/p648?a=list}{Sov. Phys. JETP {\bf 4} (1956) 648}.

\bibitem{Danilov}
G.~S.~Danilov, Zh. Eksp. Teor. Fiz. {\bf 40} (1961) 498, \href{http://www.jetp.ac.ru/cgi-bin/e/index/e/13/2/p349?a=list}{Sov. Phys.
JETP {\bf 13} (1961) 349}.

\bibitem{Faddeev-Minlos}
R.~A.~Minlos and L.~D.~Faddeev, Zh. Eksp. Teor. Fiz. {\bf 41} (1961) 1850,\\
 \href{http://www.jetp.ac.ru/cgi-bin/e/index/e/14/6/p1315?a=list}{Sov. Phys. JETP {\bf 14} (1961) 1315}.

\bibitem{Konig:2017krd}
  S.~K\"onig and D.~Lee,
  \href{http://arXiv.org/abs/1701.00279}{arXiv:1701.00279 [hep-lat]}.

\bibitem{Kaplan:1998tg}
  D.~B.~Kaplan, M.~J.~Savage and M.~B.~Wise,
  Phys.\ Lett.\ B {\bf 424} (1998) 390,
  \href{http://arXiv.org/abs/nucl-th/9801034}{arXiv:nucl-th/9801034}.
  
\bibitem{Gradstein}
I.~S.~Gradsteyn and I.~M.~Ryzhik, {\em Table of Integrals, Series, and Products,} 8th
Edition, Academic Press (2014).

\end{thebibliography}
\end{document}